# Fabrication and modeling of piezoelectric transducers for High-Frequency medical imaging

Andre-Pierre Abellard[1,2], Danjela Kuščer[1], Janez Holc[1], Franck Levassort[2], Oleksandr Noshchenko[1], Marc Lethiecq[2], Marija Kosec[1].

[1]Jožef Stefan Institute, Jamova 39, SI-1000 Ljubljana, Slovenia
[2]François-Rabelais University, UMRS Imagerie et Cerveau, Inserm U930, CNRS ERL 3106, 37032 Tours, France
andre-pierre.abellard@etu.univ-tours.fr

*Abstract* – We have studied the processing of piezoelectric thick films using electrophoretic deposition (EPD) for high-frequency ultrasound applications. Lead-zirconium-titanate (PZT) particles synthetized by solid states synthesis were dispersed in ethanol using ammonium polyacrylate (PAA). The electrophoretic deposition of PZT particles was performed at a constant-current mode. PZT thick-films deposited at 1 mA for 60 seconds were sintered at 900ºC for 2 hours in a PbO-controlled atmosphere. The scanning-electron microscopy (SEM) analysis shows that the thickness of PZT layer is uniform and that the pores are homogeneously distributed within the layer. The complex electrical impedance was measured and fitted by KLM scheme in order to deduce the dielectric, mechanical and piezoelectric parameters of the thick-films. The density and thickness of PZT thick films are used as inputs and the thickness coupling factor $k_t$, dielectric constant at constant strain and resonant frequency are deduced. The results show that homogeneous PZT thick-film structures with tailored thickness and density prepared by EPD and sintering having a resonant frequency around 20 MHz can be used for noninvasive medical ultrasound imaging and diagnostics.

## 1. INTRODUCTION

Ultrasound transducers are widely used for medical investigations and diagnostics. For imaging small objects such as skin, eye or for intracardiac and intravascular examinations, the transducer generally operates in a frequency range between 20 and 50 MHz [1, 2].

The typical transducer consists of a piezoelectric element, one or several matching layers, lenses for focusing the acoustical beam and a backing. The piezoelectric material is the most important part of the transducer since it converts electrical energy into mechanical and vice versa. Its characteristics together with its geometry contribute to many of the properties of the transducer in particular its operating frequency, sensitivity and spatial resolutions [3].

Lead-zirconate titanate (PZT) is often used as piezoelectric element due to its high piezoelectric response. For high-frequency devices, the piezoelectric element should be a few tens of micrometers thick and it can be fabricated by thick film technology such as screen-printing [4, 5]. A considerable interest is paid to geometrical focusing of the acoustic beam. In this case the piezoelectric layer is not deposited on a flat but on a curved substrate. Its shape enables focusing of the acoustic beam and therefore the lenses for focusing are not needed. This would result in a more economical and straightforward design of the transducer. Moreover, the acoustical attenuation in lenses is significant which decreases the sensitivity of the transducer.

The electrophoretic deposition (EPD) is a suitable method to realize thick-film structures on non-flat substrates [6, 7, and 8]. In this process two electrodes are immersed into a liquid containing charged ceramic particles. By applying DC electric filed, the particles move toward an electrode and deposit on it [9, 10]. To obtain the functional response of the piezoelectric material, the as-deposited layer has to be sintered at elevated temperatures [11]. The major difficulties reside in the chemical reactivity between various materials, developing of stresses in the multilayer structures and in particular in lead-containing materials and the lead-losses [11]. To process defect-free, homogeneous thick films with desired thickness and density, the properties of the suspension, the deposition process as well as the sintering procedure have to be optimized.

The substrate, used as a backing, is a porous ceramic considered as a ceramic-air composite. The air reduces the acoustical impedance of the material and increases the acoustical energy delivered to biological tissues [12].

The motivation of our work was to process piezoelectric PZT thick films with tailored thickness and density by electrophoretic deposition. The thickness and density of PZT thick films will be determined as well as the electromechanical properties of the PZT.

## 2. EXPERIMENTAL

For the experimental work we used $Pb(Zr_{0.53}, Ti_{0.47})O_3$ powder (PZT) prepared by solid state synthesis. PbO (99.9+ %, Aldrich), $ZrO_2$ (99.9 %, Tosoh) and $TiO_2$ (99.9 %, Alfa Aesar) were mixed in a planetary ball mill for 2 hours. The powder was dried and calcined two times at 950 ºC for 2 h. After the calcination the powder was milled for 8 h in an atritor mill.

PZT suspension with a solid load content of 1 vol. % was prepared by dispersing the PZT powder in an anhydrous ethanol. Ammonium polyacrylate (PAA) and n-butylamine (BA) were used as additives. We added 0.63 wt. % of both, PAA and BA, respectively, based on amount of PZT powder. The suspension was homogenized in $ZrO_2$ planetary mill for 2 hours. EPD was performed at a constant current at room temperature in a horizontal electrode cell with a platinized $Al_2O_3$ substrate that acts as a cathode and a platinized $Al_2O_3$ substrate with a PZT buffer layer that acts as an anode. The distance between the electrodes was 25 mm. The area of electrodes was 0.64 $cm^2$. The deposition was performed at a current of 1 mA for 1 min using Keithley 237 (Keithley Instruments Inc., Cleveland, USA).

After drying the layers were placed in a covered corundum vessel and sintered at 900ºC for 2 hours. Samples were sintered in the presence of 0.5 g of packing powder with composition $PbZrO_3$ with 2 wt. % of PbO. Samples are denoted as 900_2h_0.5g. Samples sintered without any packing powder are denoted as 900_2h_0g

To perform the electromechanical measurements, gold electrodes (150 nm thick with a diameter of 2 mm) were sputtered. The samples were poled at 150°C for 15 min in an oil bath at 6 kV/mm.

The X-ray powder-diffraction data was collected at room temperature using a PANalytical diffractometer (X¢Pert PRO MPD, The Netherlands) in the 2θ range from 10 to 70 degrees, by steps of 0.034 degrees, with an integration time of 100 s. The phases were identified using the PDF-2 database. The zeta-potentials of PZT particles in ethanol were measured at room temperature using a ZetaPals zeta-potential meter from Brookhaven Instruments. The conductivity of the suspension was measured at room temperature using a Cond 730 Inolab. The microstructures of samples were characterized with a scanning electron microscope (SEM) (JEOL 5800, Tokyo, Japan). The thickness of samples was measured before and after deposition by profilometer (Viking 100 Solarius Inc.). The density of the sintered thick film was obtained by a quantitative characterization of the ceramic microstructure. This characterization was obtained from SEM images turned into a binary image computerized in image-analysis software (ImageTools 3.0, University of Texas Health Science Center, San Antonio, USA) to determine the quantity of pores [13].

Dielectric, mechanical and piezoelectric parameters were deduced by measuring the complex electrical impedance around the fundamental thickness mode of resonance. The experimental set-up was composed of an HP 4395 vector analyzer and its impedance test kit. To simulate the theoretical behavior of the electrical impedance of the samples as a function of frequency for the thickness mode, an equivalent electrical circuit model was used. In this paper, the KLM scheme [14] was retained and a fitting process and used to deduce from the experimental data the thickness-mode parameters of the EPD piezoelectric thick films. The structure of the samples is composed of four inert layers and one piezoelectric layer (Figure 1).

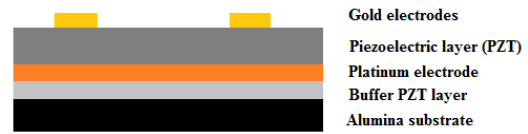

Figure 1: Schematic representation of thick-film structure.

The four inert layers are an alumina substrate, an unpoled PZT buffer layer and two electrodes, the parameters of those layers are fed in the KLM model and considered as constant. The determination and the corresponding accuracy of the parameters are of first priority [15, 16]. These materials and corresponding data base are well known except for porous PZT. For this last material, a homogenization model was used [17, 18] (according to the porosity content) to deduce the longitudinal wave velocity and attenuation coefficient. Finally, with the KLM model, five parameters were deduced (i.e. the thickness mode coupling factor $k_t$, the longitudinal wave velocity $v_l$, the dielectric constant at constant strain $\varepsilon_{33}^S$, and the loss factor (mechanical and dielectric). On these basis and using additional measurements such as the density or the thickness of the piezoelectric thick film, acoustical impedance (Z) as well as the resonant frequency of the thick film ($f_0$) in free mechanical resonance condition were deduced. Accuracy of this characterization method for a multilayer structure was discussed in previous work [19, 20].

## 3. RESULTS AND DISCUSSION

The X-ray powder diffraction pattern of PZT powder after calcination is shown in Figure 2. All diffraction peaks correspond to tetragonal (PDF 00-050-0346) and rhombohedral (PDF 01-073-2022) perovskite PZT structure [21].

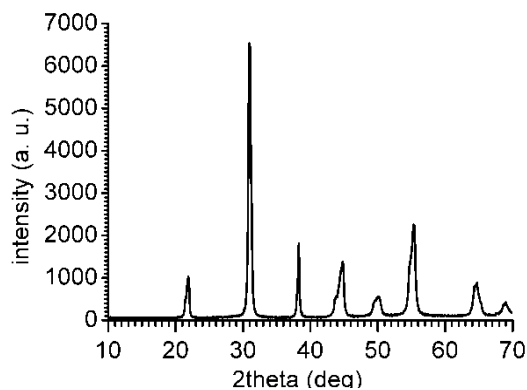

Figure 2: X-ray diffraction of PZT powder calcined at 950°C for 2 hours.

We prepared an ethanol-based PZT suspension with a solid load of 1 vol. % of PZT. The zeta potential of PZT particles of -62 ± 5 mV indicates stable suspension. The conductivity of the suspension was 16 µS/cm at 23°C.

PZT particles were deposited on the substrates at a constant current of 1 mA for 1 min. We obtain a crack-free, uniform deposit as shown in Figure 3. The thickness of the deposit was 70 ± 5 µm and a density 50 ± 5 % of the theoretical density of PZT determined from dimensions and mass of the PZT layer.

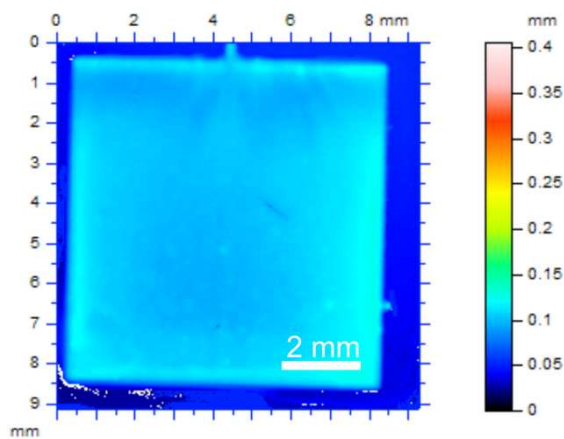

Figure 3: Top surface of PZT layer deposited at 1 mA for 60 seconds.

The PZT deposits were sintered at 900 °C for 2 hours. The microstructures of the samples sintered without packing powder (denoted as 900_2h_0g) and with 0.5 g of packing powder (denoted as 900_2h_0.5g) are shown in Figure 4 and Figure 5, respectively. From the images it is evident that the thickness of the piezoelectric layer is uniform. The layer is porous and the pores are homogenously distributed in the PZT matrix. The PZT layer of the sample 900_2h_0g had a thickness of 55 ± 1 µm and a density of 69 ± 1 %. PZT particles are small and not connected as it is evident in Figure 4b.

The PZT layer of sample 900_2h_0.5g is slightly thinner, i.e. 50 ± 1 µm. The density of the PZT layer is 72 ± 2 %. PZT particles are connected as it is evident from Figure 5b. The results illustrate that the sintering of PZT thick film in the presence of the lead-rich atmosphere results in a thinner and denser PZT layer.

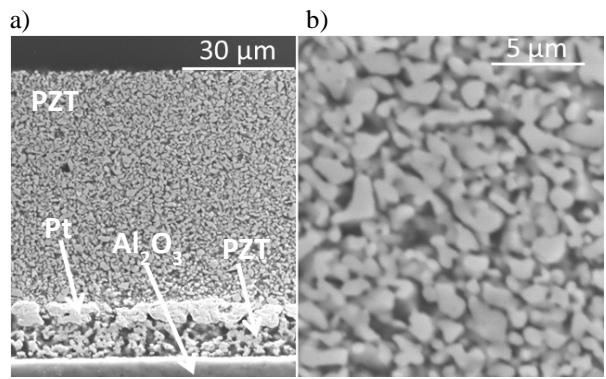

Figure 4: PZT sample sintered without packing powder (900_2h_0g). a) Cross section of PZT thick-film deposited on Pt/PZT/Al$_2$O$_3$ substrate; b) PZT layer.

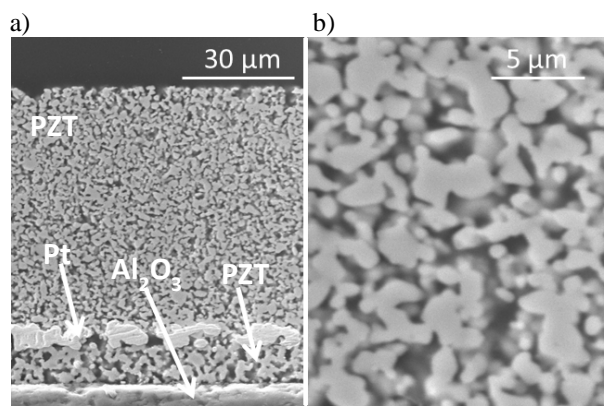

Figure 5: PZT sample sintered with 0.5 g of packing powder (900_2h_0.5g). a) Cross section of PZT thick-film deposited on Pt/PZT/Al$_2$O$_3$ substrate; b) PZT layer.

The complex electrical impedance of sample 900_2h_0g and the sample 900_2h_0.5g are shown in Figure 6 and in Figure 7, respectively.

By using the KLM scheme and a fitting process, we deduced two parameters, the thickness coupling factor $k_t$ and the dielectric constant at constant strain $\varepsilon_{33}^S$. We determined the resonance frequency of the PZT piezoelectric layer from complex electrical impedance measurement represented by a dashed gray line. For sample 900_2h_0g, a coupling factor of 20%, an apparent dielectric constant at constant strain of 59 was measured as well as a resonance frequency of 22.9 MHz. For Sample 900_2h_0.5g, a coupling factor of 33%, an apparent dielectric constant at constant strain of 94 and a resonance frequency of 18.1 MHz were found.

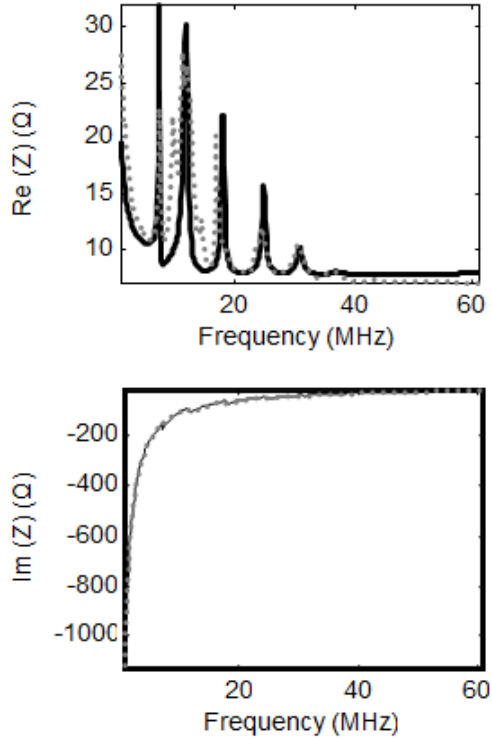

Figure 6: Complex electrical impedance of sample 900_2h_0g (solid black line: theoretical, dashed gray line: experimental).

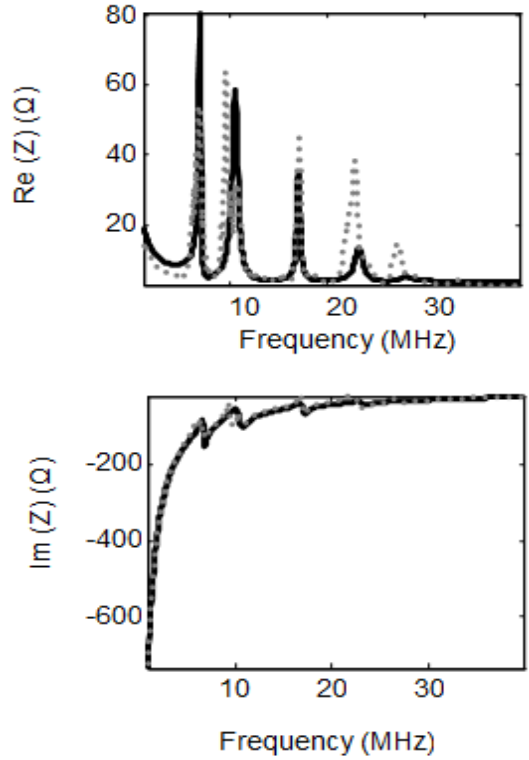

Figure 7: complex electrical impedance of sample 900_2h_0.5g (solid black line: theoretical, dashed gray line: experimental).

The apparent dielectric constant and the thickness coupling factor of sample 900_2h_0.5g are higher than those of the sample 900_2h_0g. This is consistent with the microstructure of the thick films. The PZT layer is denser and the particles are connected together in sample 900_2h_0.5g. This probably explains the higher $k_t$ and higher apparent dielectric constant. Our results show that the sintering of PZT thick-film structures in a lead-rich atmosphere improves the electromechanical performance of the PZT piezoelectric layer.

The resonance frequency of PZT thick films, which depends on the PZT film thickness, is around 20 MHz. This is in the range of frequencies for high resolution medical imaging.

## 4. CONCLUSIONS

We prepared homogenous piezoelectric PZT thick-film structures with uniform thickness for ultrasound transducer applications. The PZT powder prepared by solid state synthesis was dispersed in ethanol-based suspension and deposited on electroded alumina substrate by electrophoretic deposition. The density of the piezoelectric layer was controlled by sintering atmosphere. The PZT thick film with a thickness of

50 μm and a density of 72 % of the theoretical value has a thickness coupling factor of 33 % and an apparent dielectric constant at a constant strain of 94. The resonance frequency of the thick-film structures are around 20 MHz and can be used for medical investigation and diagnosis.

Our future work, we will be oriented toward processing of PZT thick-film structures with a resonance frequency of 50 MHz. We will increase the resonant frequency of the transducer by controlling the thickness and density of PZT layer. We will tailor the properties of the suspension, the deposition conditions during EPD process and the sintering process by varying temperature, time and atmosphere.

**Acknowledgments**

Mr. Silvo Drnovšek, Mr. Tadej Rojac, Mrs. Alja Kupec, Mrs. Milena Pajić, Mrs. Petra Kuzman and Mrs. Jena Cilenšek are acknowledged for technical assistance.

**References**

[1] T.R. Gururaja, "Piezoelectrics for medical ultrasonic imaging", Am. Ceram. Soc. Bull., vol. 73, no. 5, pp. 50-55, 1994.
[2] K.K. Shung, J.M. Cannata, Q.F. Zhou, "Piezoelectric materials for high frequency medical imaging applications: A review", J. Electroceram., vol. 19, no. 1, pp. 139-145, 2007.
[3] M. Lethiecq, F. Levassort, D. Certon, L.P. Tran-Huu-Hue, "Piezoelectric and acoustic materials for transducer applications", Chapter 10: "Piezoelectric transducer design for ultrasonic medical diagnosis & NDE", ed. A. Safari, E.K. Akdogan, Springer, New York, pp. 191-215, 2008.
[4] P. Maréchal, F. Levassort, J. Holc, L.P. Tran-Huu-Hue, M. Kosec, M. Lethiecq, "High frequency transducers based on integrated piezoelectric thick films for medical imaging", IEEE TUFFC, vol. 53, no. 8, pp. 1524-1533, 2006.
[5] P. Tran-Huu-Hue, F. Levassort, F. Vander Meulen, J. Holc, M. Kosec, M. Lethiecq, "Preparation and electromechanical properties of PZT/PGO thick films on alumina substrate", J. Eur. Ceram. Soc., vol. 21, no. 10-11, pp. 1445-1449, 2000.
[6] P. Sarkar, P.S. Nicholson, "Electrophoretic deposition (EPD): Mechanisms, kinetics and applications to ceramics", J. Am. Ceram. Soc., vol. 79, no. 8, pp. 1987-2022, 1996.
[7] I. Corni, M.P. Ryan, A.R. Boccaccini, "Electrophoretic deposition: From traditional ceramics to nanotechnology", J. Eur. Ceram. Soc., vol. 28, no. 7, pp. 1353-1367, 2008.
[8] B. Ferrari, R. Moreno, "EPD kinetics: A review", J. Eur. Ceram. Soc., vol. 30, no. 5, pp. 1069-1078, 2010.
[9] J. Ma, W. Cheng, "Electrophoretic deposition of lead zirconate titanate ceramics", J. Am. Ceram. Soc., vol. 85, no. 7, pp. 1735-1737, 2002.
[10] D. Kuščer, M. Kosec, "Electrophoretic deposition of lead-zirconate-titanate perovskite thick films with low sintering temperature", Key Eng. Mater., vol. 412, pp. 101-106, 2009.
[11] L. Pardo, J. Ricote, M. Kosec, D. Kuscer, J. Holc, "Multifunctional Polycrystalline Ferroelectric Materials: Processing and Properties", Chapter 2: "Processing of Ferroelectric Ceramic Thick Films", Springer Series in materials, Springer Netherlands, vol. 140, pp. 39-61, 2011.
[12] C. Galassi, "Processing of porous ceramics: Piezoelectric materials", J. Eur. Ceram. Soc., vol. 26, no. 14, pp. 2951-2958, 2006.
[13] D. Kuscer, J. Korzekwa, M. Kosec, R. Skulski, "A- and B-compensated PLZT x/90/10: Sintering and microstructural analysis", J. Eur. Ceram. Soc., vol. 27, no. 6, pp. 4499-4507, 2007.
[14] D. A. Leedom, R. Kimholtz, and G. L. Matthaei, "New equivalent circuit for elementary piezoelectric transducers", Electronic Lett., vol. 6, no. 13, pp. 398-399, 1971.
[15] A.R. Selfridge, "Approximate material properties in isotropic materials", IEEE Trans. Sonics Ultrason., vol. 32, no. 3, pp. 381-394, 1985.
[16] http://www.ondacorp.com/tecref_acoustictable.shtml (Last viewed 23th of May 2010)
[17] F. Levassort, M. Lethiecq, D. Certon, F. Patat, "A matrix method for modeling electroelastic moduli of 0-3 piezo-composites", IEEE TUFFC, vol. 44, no. 2, pp. 445-452, 1997.
[18] F. Levassort, J. Holc, E. Ringgaard, T. Bove, M. Kosec, M. Lethiecq, "Fabrication, modeling and use of porous ceramics for ultrasonic transducer applications", J. Electroceram., vol. 19, no. 1, pp. 125-137, 2007.
[19] M. Lukacs, T. Olding, M. Sayer, R. Tasker, S. Sherrit, "Thickness mode material constants of a supported piezoelectric film", J. Appl. Phys., vol. 85, no. 5, pp. 2835-2843, 1999.
[20] A. Bardaine, P. Boy, P. Belleville, O. Acher, F. Levassort, "Improvement of composite sol-gel process for manufacturing 40 μm piezoelectric thick films", J. Eur. Ceram. Soc., vol. 28, no. 8, 1649-1655, 2008.
[21] PDF-ICDD, PCPDFWin Version 2.2, June 2001, International Centre for Diffraction Data, 2002.